\documentclass{ws-ijmpd}

\newcommand{\ket}[1]{|#1\rangle}    %ket
\begin{document}

\markboth{R. Gambini, L. P. Garc\'{\i}a-Pintos, J. Pullin}
{Undecidability as solution to the problem of
  measurement}

\catchline{}{}{}{}{}
%
%%%%%%%%%%%%%%%%%%%%%%%%%%%%%%%%%%%%%%%%%%%%%%%%%%%%%%%%%%%%%%%%%%%%

\title{Undecidability as solution to the problem of
  measurement: fundamental criterion for the production of events}

\author{Rodolfo Gambini, Luis Pedro Garc\'{\i}a-Pintos}
\address {Instituto de F\'{\i}sica,
Facultad de Ciencias, Igu\'a 4225, esq. Mataojo, Montevideo, Uruguay.}

\author{Jorge Pullin}
\address{Department of Physics and Astronomy, Louisiana State
University\\ Baton Rouge, LA 70803-4001}

\maketitle

\begin{history}
\received{August 4 2010}
\revised{Day Month Year}
\comby{Managing Editor}
\end{history}

\begin{abstract}
  In recent papers we put forth a new interpretation of quantum
  mechanics, colloquially known as ``the Montevideo interpretation''.
  This interpretation is based on taking into account fundamental
  limits that gravity imposes on the measurement process. As a
  consequence one has that situations develop where a reduction
  process is undecidable from an evolution operator. When such a
  situation is achieved, an event has taken place. In this paper we
  sharpen the definition of when and how events occur, more precisely
  we give sufficient conditions for the occurrence of events. We probe
  the new definition in an example. In particular we show that the
  concept of undecidability used is not ``FAPP'' (for all practical
  purposes), but fundamental.
\end{abstract}

\section{Introduction}

The problem of measurement in quantum mechanics arises in standard
treatments as the requirement of an external process called the
reduction process when a measurement takes place. Such process is not
contained within the unitary evolution of the quantum theory but has
to be postulated externally and is not unitary. It is usually
justified through the interaction with a large, classical measuring
device and an environment with many degrees of freedom. Attempts to
formulate a coherent framework with purely quantum rules taking into
account the environment have however failed to provide a consistent
picture of quantum mechanics and the measurement process. Objections
have been levied onto two aspects of the solution of the problem of
measurement through decoherence. First of all, although a quantum
system interacting with an environment with many degrees of freedom
will very likely give the appearance that the initial quantum
coherence of the system is lost, since the evolution of the system
plus environment is unitary, that coherence could potentially be
regained. This phenomenon is called ``revival'', and although in
practice it may take a very long time to arise, it exists as an issue
of principle.  It is always in principle possible to recover the
information lost during the measurement process carrying out a
measurement that includes the environment. The fact that such
measurements are hard to carry out in practice does not prevent the
issue from existing as a conceptual problem. The second criticism has
to do with the fact that in a picture where evolution is unitary
``nothing ever occurs''. That is, one may have a way of characterizing
the interaction with the environment that yields a density matrix with
zero off diagonal elements, but one is left with a description of a
set of coexistent options and not with a definite assignment of
probabilities to alternative options, as one has after a measurement
has taken place.

We have recently proposed a solution to the problem of measurement,
leading to a new interpretation of quantum mechanics commonly known as
``the Montevideo interpretation'' \cite{montevideo}. The idea is that gravity
fundamentally limits how accurate our measurements of space and time
can be. This requires reformulating quantum mechanics in terms of real
clocks and rods, that have errors in their measurements. The resulting
picture of quantum theory is one where there is fundamental loss of
quantum coherence: pure states evolve into mixed states. This
eliminates the problem of revivals, just waiting longer does not
improve things as more quantum coherence is lost. One also has an
operational definition for when an event takes place: when the
fundamental loss of coherence is such that one cannot distinguish
the unitary evolution from a reduction, an event has taken place.
We call this situation ``undecidability'' between reduction and
unitary evolution.

A criticism that could be levied against our proposal is that there
was no clear criterion given for when undecidability takes place. In
particular, is the criterion supplied fundamental or is it ``FAPP''
(for all practical purposes). To analyze this in detail we will
consider a model where the quantum system, the measuring apparatus and
the environment are under control. This is the case of an example that
we have already considered in this context and is a variation of a
model proposed by Zurek in the context of decoherence. In section 2 we
describe the model briefly, in section 3 we discuss undecidability and
in section 4 we discuss a fundamental limit in the measurement of
spins we will need to discuss the example. In section 5 we outline the
sharp criterion for a production of an event and illustrate it with an
example. We end with a discussion.

\section{The model}

In a previous paper\cite{undecidability} we introduced a model of
decoherence in order to study the appearance of undecidability and its
implications for the measurement problem. The model is a variation of
a model presented by Zurek\cite{zurek}. Here we outline some of the
results in order to make this paper self contained.

The model consists of a spin $S$ located in the center of a chamber with
a magnetic field $B$ pointed in the $z$ direction. Into the chamber flow,
one by one, a set of $N$ ``environmental'' spins. The interaction
Hamiltonian for the $k$-th spin of the environment is,
\begin{equation}
\hat{H}_k = \hat{H}^B_k + \hat{H}^{\rm int}_k,
\end{equation}
with,
\begin{equation}
\hat{H}^B_k = \gamma_1 B \hat{S}_z\otimes \hat{I} _k +\gamma_2
B\hat{I}\otimes S^{k}_z,
\end{equation}
and
\begin{equation}
\hat{H}^{\rm int}_k = f_k \left(
\hat{S}_x \hat{S}_x^k +
\hat{S}_y \hat{S}_y^k +
\hat{S}_z \hat{S}_z^k\right),
\end{equation}
where $\gamma_1$ y $\gamma_2$ are the magnetic moments of the central
and environment spins respectively and the $\hat{S}$ are spin operators.

Starting with the initial state,
\begin{equation}
  \ket{\Psi} = (a\ket{\uparrow} + b\ket{\downarrow}) \bigotimes_k^N
(\alpha_k \ket{\uparrow}_k + \beta_k \ket{\downarrow}_k) \label{inicial}
\end{equation}
one can see that in the limit of weak couplings ($f_k \ll
{{B\gamma_1,B\gamma_2}}$) one has decoherence in the $z$ basis,
yielding a state after the passage of $N$ spins,

\begin{eqnarray}
  \vert\Psi(t)\rangle
  &=& a \ket{\uparrow} \prod_{k=1}^N\otimes\left[
    \alpha_k\exp\left(i\int_0^\tau dt f_k \right)\ket{\uparrow}_k
    + \beta_k \exp\left(-i\int_0^\tau dt f_k \right)\ket{\downarrow}_k\right]
\label{estado} \\
  &&+ b \ket{\downarrow}
  \prod_{k=1}^N\otimes\left[
    \alpha_k\exp\left(-i\int_0^\tau dt f_k \right)\ket{\uparrow}_k
    + \beta_k \exp\left(i\int_0^\tau dt f_k \right)\ket{\downarrow}_k\right].
\nonumber
\end{eqnarray}
where $\tau$ is the time of flight of the environment spins passing
through the chamber. The reduced density matrix becomes,
\begin{equation}
\hat{\rho}_S=\left(
\begin{array}{cc}
\vert a\vert ^2 & ab^*\\
a^*b & \vert b\vert ^2
\end{array}\right)
\begin{array}{c}
\longrightarrow\\
{\scriptstyle N\gg 1}
\end{array}
\left(
\begin{array}{cc}
\vert a\vert ^2 & \sim 0\\
\sim 0& \vert b\vert^2
\end{array}
\right).
\end{equation}

This implies that from the point of view of local observables the
system behaves \emph{almost} as if it were in one of the possible
states and no local experiment allows to check if it is in a quantum
superposition or not.  With those types of measurements it therefore
becomes increasingly more difficult to check if evolution was unitary
or a collapse of the wavefunction has taken place. This fact is
sometimes used to argue that this effect provides a solution to the
measurement problem.  However, as noted in \cite{undecidability},
there exists observables of global nature, for instance one proposed
by d'Espagnat \cite{despagnat}, whose expectation value is different
depending on if collapse has or has not taken place. It is given by,
\begin{equation}
\hat{M} \equiv \hat{S}_x \otimes\prod_k^N \hat{S}^k_x.
\end{equation}
One has that $\langle \hat{M} \rangle_{collapse} = 0$ whereas,
\begin{equation}
\langle \psi\vert M\vert\psi\rangle =
ab^*\prod_k^N\left[\alpha_k\beta_k^* +\alpha^*_k\beta_k\right]
e^{-2i\Omega_k\tau} + a^*b
\prod_k^N\left[\alpha_k\beta_k^* +\alpha^*_k\beta_k\right]
e^{2i\Omega_k\tau} \ne 0,
\end{equation}
with $\Omega_k \equiv \sqrt{4f_k^2+B^2(\gamma_1-\gamma_2)^2}$ and
$\tau$ is the time of flight of the environmental spins through the chamber.
One therefore has the possibility of determining experimentally if a
collapse has taken place or if the system remains in a superposition
that behaves classically only when probed with local observables.

\section{Undecidability and the FAPP problem}

As we discussed in the introduction, the notion of undecidability
arises due to the fundamental limitations in the measurement of times
and lengths that gravity imposes on us. When one takes into account
that the Newtonian time $t$ that appears in the Schr\"odinger equation
is really not observable (it is an ideally classical quantity whereas
all real clocks are quantum systems and the best one can do it to
associate the eigenvalue $T$ of some quantum operators $\hat{T}$ to
the measurement of time), one notes that the resulting evolution loses
quantum coherence\cite{obregon}. The rate at which coherence is lost
depends on how good or bad the clock is. Using heuristic
estimates\cite{ng} for the fundamental uncertainty of a clock one
notes that off diagonal elements of the density matrices die off at a
rate, $\exp \Big(-\frac{2}{3} \omega_{nm}^2 T_{\rm P}^{4/3} T^{2/3}
\Big)$, where $\omega_{nm}$ is the Bohr frequency between levels $n$
and $m$ of an energy eigenbasis, $T$ is the time measured by the clock
and $T_{\rm P}\sim 10^{-44}$s is Planck's time.

Taking this effect into account the expectation value of the
observable $\hat{M}$ is,
\begin{eqnarray}
\langle \hat{M}\rangle &=&
a b^* e^{-i2 N \Omega T} e^{-4 N B^2 (\gamma_1-\gamma_2)^2 \theta}
\prod_{k}^N \left[ \alpha_k \beta_k^* e^{-16 B^2\gamma_1\gamma_2\theta}
+\alpha_k^* \beta_k\right]\\
&&+b a^*
e^{i2 N \Omega T} e^{-4 N B^2 (\gamma_1-\gamma_2)^2 \theta}
\prod_{k}^N \left[ \alpha_k \beta_k^*
+\alpha_k^* \beta_ke^{-16 B^2\gamma_1\gamma_2\theta}\right]
\end{eqnarray}
where $\Omega \equiv B(\gamma_1 -\gamma_2)$, $\theta \equiv
\frac{3}{2} T_{\rm P}^{4/3} \tau^{2/3}$, $\tau$ is the time of flight
of the environment spins within the chamber and $T$ is the length of
the experiment.

There exist a series of conditions for the experiment to be feasible
that imply certain inequalities,
\begin{eqnarray}
&&a) \qquad 1< f\tau =\frac{\mu\gamma_1 \gamma_2}{\hbar}\frac{\tau}{d^3},
\label{a}\\
&&b) \qquad \Delta x \sim \sqrt{\frac{\hbar T}{m}}, \label{b}\\
&&c)\qquad f\ll \vert B(\gamma_1 -\gamma_2)\vert, \label{c}\\
&&d)\qquad \langle \hat{M} \rangle \sim
\exp\left(-6 N B^2(\gamma_1-\gamma_2)^2
T_{\rm Planck}^{4/3} \tau^{2/3}\right),\label{d}
\end{eqnarray}
with $f$ the interaction energy between spins which was assumed
constant through the cell, $\mu$ the permeability of the vacuum, $d$
the impact parameter of the spins of the environment, $m$ their mass,
and $\Delta x$ the spatial extent of the environment particles.

Condition a) stems from ensuring that the coupling between spins is
not too weak, in order for decoherence to occur; b) is to prevent the
particles of the environment from dispersing too much and therefore
making us unable to find them within the detectors at the end of the
experiment; c) is the condition for decoherence to be in the $z$
basis, as we mentioned; d) is an estimation of the
the expectation value of the observable when the effect of
the real clock is taken into
account. For details of the derivation of these conditions see
our previous paper\cite{undecidability}.

As can be seen, there is an exponentially decreasing factor that makes
the expectation value of the observable tend to that of the case in
which collapse occurs as $\tau$ and $N$ increase. One has that,
\begin{equation}
\langle \hat{M} \rangle \approx \langle \hat{M} \rangle_{collapse}.
\end{equation}

>From the previous discussion one can gather that as one considers
environments with a larger number of degrees of freedom and as longer
time measurements are considered, distinguishing between collapse and
unitary evolution becomes harder.  But can this be considered a
fundamental claim?  Could one not, repeating the experiment many
times, distinguish one case from the other? Is such a construction
only a solution ``for all practical purposes'' (FAPP)?  A similar
criticism could be levied in interpretations based on environmental
decoherence, even ignoring the problem of ``revivals'' or of potential
global observables that distinguish both cases. Since environmental
decoherence effects make the off-diagonal elements of the density
matrix small but non-zero, one could make the coherences apparent by
repeating the experiment in question a large number of times.

In the next sections we will show that there exist fundamental
uncertainties in the measurement of quantum observables that prevent
one from distinguishing the presence of small values in the density
matrix with the elements vanishing.

\section{Fundamental limit on the measurement of spins}

Following Brukner and Kofler\cite{koflerbrukner}, let us consider a
device for measuring the spin in a given direction (for instance a
Stern--Gerlach setup). If $L$ is the angular momentum of the device
and $\theta$ the angle that indicates the direction it is measuring,
the uncertainty principle implies that\footnote{For small angular
uncertainties. },
\begin{equation}
\Delta L \Delta \theta \ge \frac{\hbar}{2}.
\end{equation}
The commutator between the angle and angular momentum operator is,
\begin{equation}
[\theta,L] = i\hbar
\end{equation}
whereas the Hamiltonian of the measuring device may be modeled by a
rigid rotator,
\begin{equation}
H = \frac{L^2}{2I},
\end{equation}
where $I \approx MR^2$ is its moment of inertia, $M$ its mass and $R$
its characteristic length.

The evolution equation for the operator $\theta(t)$ in Heisenberg's
representation is
\begin{equation}
\frac{d \theta}{dt} = -i \frac{[\theta,H]}{\hbar} = \frac{L}{I},
\end{equation}
and,
\begin{equation}
\theta(\tau) = \theta(0) + \frac{L\tau}{I}.
\end{equation}

Recalling that for two observables $A$ and $B$ we have that $\Delta A
\Delta B \ge 1/2 \vert \langle [A,B] \rangle \vert$, we obtain,
\begin{equation}
\Delta \theta (0) \Delta \theta (\tau) \ge \frac{\hbar \tau}{2I}
\approx \frac{\hbar \tau}{MR^2}.
\end{equation}
Therefore at least one of the observables $\theta(0)$ and
$\theta(\tau)$ has a dispersion such that,
\begin{equation}
\Delta \theta \gtrsim \frac{1}{R} \sqrt{\frac{\hbar \tau}{M}}.
\end{equation}

The parameters of the measuring device cannot take any given
value. Special relativity bounds the characteristic size $R$, which
cannot exceed the length light would travel in the time it takes the
measurement, therefore $R \le c \tau$, from where
\begin{equation}
\Delta \theta \gtrsim \sqrt{\frac{\hbar}{cMR}}.
\end{equation}
General relativity adds another constraint, $R \ge 2GM/c^2$, since $R$
must be bigger than the Schwarzschild radius associated with the mass
$M$. With this, we get,
\begin{equation}
\Delta \theta \gtrsim \frac{l_P}{R},
\end{equation}
where $l_P \equiv \sqrt{\hbar G/c^3} \approx 10^{-35}m$. If we take
the radius of the observable universe as a characteristic length, $R \approx
10^{27}m$,
we reach a fundamental bound on the measurement of the angle,
\begin{equation}
\Delta \theta \ge 10^{-62}.
\end{equation}

So we see that from a very general quantum mechanical analysis
together with bounds from special and general relativity we have a
fundamental uncertainty in the measurements of angles. Let us see what
consequences follow if we wish to measure the expectation value of the
spin in the $z$ direction. If the spin's state is
$\Phi = a\ket{\uparrow} + b\ket{\downarrow}$, then,
\begin{equation}
\langle S_z \rangle = \vert a \vert^2 - \vert b \vert^2.
\end{equation}
So instead of measuring $S_z$ we really are measuring
$S_z^{\Delta \theta} = S.\hat{n}$, con $\hat{n} =
(\sin(\Delta \theta), 0, \cos(\Delta \theta))$. Then,
\begin{equation}
\langle S_z^{\Delta \theta} \rangle = \sin(\Delta \theta)
\langle S_x \rangle + \cos(\Delta \theta) \langle S_z
\rangle \approx \langle S_z \rangle + \Delta \theta (ab^* + a^*b)
- \frac{(\Delta \theta)^2}{2} \langle S_z \rangle. \label{valoresp}
\end{equation}
We therefore see that one has an error of order\footnote{Up to now we
  are ignoring errors associated with the preparation of the initial
  state.}  $\Delta \theta$
(or $(\Delta \theta)^2$ depending on the initial state). It is
worthwhile pondering why these angular errors do not average to zero
since when one measures $\langle S_z \rangle$ one carries out the
experiment several times. The point to emphasize is that the
uncertainty we are talking about is different from that stemming from
a procedure where there are random measurement errors. In such a case
one would have a certain probability that the device measures a
direction different from the desired one. In our case the experimental
device has an intrinsic error $\Delta \theta$ which implies that one
{\em cannot know} in which direction the measurement took place. An
analogy would be the random errors and the perception errors in a
measurement. Random errors can be minimized measuring many
times. Perception errors cannot.

\section{Solving the FAPP problem}

Let us apply the result of the previous section to the observable $M$
we discussed in sections 2 and 3. We saw that its expectation value was,
(\ref{d}),
\begin{equation}
\langle \hat{M} \rangle \sim \exp\left(-6 N B^2(\gamma_1-\gamma_2)^2
T_{\rm Planck}^{4/3} \tau^{2/3}\right) \equiv e^{-K}.
\end{equation}
Let us recall that to distinguish if there is collapse or not, one
needs to distinguish $\langle \hat{M} \rangle$ from $0$. However,
according to the result of the previous section the observable will
have an error that depends on $\Delta \theta$. If this error is larger than
$\langle \hat{M} \rangle$, there would be no way of distinguishing
collapse from a unitary evolution.

To compute the error let us recall the expression for the observable
and lets add the uncertainty in the direction that is measured,
\begin{equation}
\hat{M}^{\Delta \theta} \equiv \hat{S}_x^{\Delta \theta} \otimes\prod_k^N \hat{S}^{k,\Delta \theta}_x,
\end{equation}
and,
\begin{equation}
\hat{S}_x^{\Delta \theta} \approx \hat{S}_x + \Delta \theta \hat{S}_z.
\end{equation}

Therefore, the observable we really measure will have a term
$\hat{S}_x \otimes\prod_k^N \hat{S}^k_x$, a term $(\Delta
\theta)^{N+1} \hat{S}_z \otimes\prod_k^N \hat{S}^k_z$, and cross terms
of the form $(\Delta \theta)^{n} \otimes\prod_i^{N-n} \hat{S}^i_x
\otimes\prod_j^n \hat{S}^j_z$. Each of the new terms containing powers
of $\Delta \theta$ will add noise to the measurement of the
observable. Let us define $E(\Delta \theta)$ as all the terms except
$\hat{S}_x \otimes\prod_k^N \hat{S}^k_x$ and $(\Delta \theta)^{N+1}
\hat{S}_z \otimes\prod_k^N \hat{S}^k_z$. We then get
\begin{equation}
  \hat{M}^{\Delta \theta} \approx \hat{S}_x \otimes\prod_k^N
\hat{S}^k_x + (\Delta \theta)^{N+1} \hat{S}_z \otimes\prod_k^N
\hat{S}^k_z + E(\Delta \theta) = \hat{M} + (\Delta \theta)^{N+1}
\hat{S}_z \otimes\prod_k^N \hat{S}^k_z + E(\Delta \theta).
\end{equation}
Using (\ref{inicial}) as initial state we have,
\begin{equation}
\langle \hat{M}^{\Delta \theta} \rangle \sim e^{-K}
+ (\Delta \theta)^N (\vert a \vert^2
- \vert b \vert^2) \prod_k^N (\vert \alpha_k \vert^2
- \vert \beta_k \vert^2) + \langle E(\Delta \theta) \rangle . \label{observable}
\end{equation}

To demonstrate the occurrence of undecidability it is enough to focus
on the error that goes like $(\Delta \theta)^N$, the terms in $\langle
E(\Delta \theta) \rangle$ will only add further noise to the
measurement. As can be seen, it could still happen, depending on the
initial state, that the error we are concentrating on vanishes. In
fact, in our previous paper\cite{undecidability} we saw that the state
that is convenient for the experiment would be such that $\vert
\alpha_k \vert^2 = \vert \beta_k \vert^2$. However, given the errors
in measuring the components of the spin, it is also impossible to
prepare perfect initial states. For instance instead of preparing
$\alpha_k \ket{\uparrow}_k + \beta_k \ket{\downarrow}_k$ one would
prepare $\alpha_k \ket{\uparrow}_k^{\Delta \theta} + \beta_k
\ket{\downarrow}_k^{\Delta \theta}$, with
$\big(\ket{\uparrow}_k^{\Delta \theta},\ket{\downarrow}_k^{\Delta
  \theta}\big)$ eigenstates of the observable $S_z + \Delta \theta
S_x$ (to first order in $\Delta \theta$). These are given by,
\begin{equation}
\ket{\uparrow}_k^{\Delta \theta} = \ket{\uparrow}_k
+ \frac{\Delta \theta}{2}\ket{\downarrow}_k,
\qquad \ket{\downarrow}_k^{\Delta \theta}
= \ket{\downarrow}_k - \frac{\Delta \theta}{2}\ket{\uparrow}_k
\end{equation}
Therefore the prepared state is,
\begin{equation}
  \alpha_k \ket{\uparrow}_k^{\Delta \theta} + 
\beta_k \ket{\downarrow}_k^{\Delta \theta} = 
\Big(\alpha_k - \frac{\Delta \theta}{2}\beta_k \Big) 
\ket{\uparrow}_k + 
\Big(\beta_k + \frac{\Delta \theta}{2}\alpha_k \Big) 
\ket{\downarrow}_k. \label{probs}
\end{equation}
As can be seen the probability amplitudes are not exactly the wanted
ones. Using these in (\ref{observable}) one gets,
\begin{eqnarray}
&&\langle \hat{M}^{\Delta \theta} \rangle \sim e^{-K}
+ \langle E(\Delta \theta) \rangle\\
&&+(\Delta \theta)^N \Big(\vert a \vert^2
- \vert b \vert^2 + \Delta \theta (ab^* + a^*b)\Big)
\prod_k^N \Big(\vert \alpha_k \vert^2
- \vert \beta_k \vert^2 + \Delta \theta
(\alpha_k \beta_k^* + \alpha_k^* \beta_k)\Big) .\nonumber
\end{eqnarray}
Even if one wishes to impose the optimal condition
$\vert \alpha_k \vert^2 = \vert \beta_k \vert^2$, the preparation
errors will lead to the observable $\hat{M}$ having an associated
error of the order of  $(\Delta \theta)^{2N}$,
\begin{equation}
\langle \hat{M}^{\Delta \theta} \rangle \sim e^{-K} \pm 
(\Delta \theta)^{2N} + \langle E(\Delta \theta) \rangle.
\end{equation}

Using the conditions (\ref{a})-(\ref{d}) one finds out that $K$ satisfies,
\begin{equation}
K \gg \frac{N^5 T_{\rm Planck}^{4/3} \hbar^{20/3}}{m^4
(\gamma_1 \gamma_2)^{8/3} \mu^{8/3}}.
\end{equation}

So we have
\begin{equation}
e^{-K} \ll \exp \bigg[- \frac{N^5 T_{\rm Planck}^{4/3} \hbar^{20/3}}{m^4 (\gamma_1 \gamma_2)^{8/3} \mu^{8/3}} \bigg]
\end{equation}
and given the strong dependence of $e^{-K}$ on $N$, it follows
that\footnote{The behavior of $K$ as $N^5$ was reached by using
  inequalities (\ref{a})-(\ref{c}) taking into consideration all
  aspects of this particular model, so it need not be the same for
  other models.}
\begin{equation}
e^{-K} \le (\Delta \theta)^{2N} + \langle E(\Delta \theta) \rangle,
\end{equation}
with which the result that one obtains with or without collapse differ
less than the observable $\hat{M}$ error and therefore it is
impossible to distinguish both cases experimentally.

Let us finally see how these considerations also make it impossible to
check if the system is or not in a quantum superposition through local
observables (as we discussed in section 3 the fundamental decoherence
effect does not eliminate entirely the interference terms). From
equation (\ref{estado}) for the final state of the system, we see that
the decoherence factor that multiplies the interference term in the
reduced density matrix is,
\begin{equation}
z \equiv \prod_k^N \Bigg[ \cos\bigg(2\int_0^\tau dt f_k \bigg)
+ i\big(\vert \alpha_k \vert^2 - \vert \beta_k \vert^2\big)
\sin\bigg(2\int_0^\tau dt f_k \bigg)  \Bigg].
\end{equation}
Let us consider the ideal situation for the experiment where
$\vert \alpha_k \vert^2 - \vert \beta_k \vert^2 = 0$. We have,
\begin{equation}
z \sim \cos\bigg(2\int_0^\tau dt f_k \bigg)^N.
\end{equation}
This factor $z$ would appear on measurements of the observable $S_x$, as,
\begin{equation}
\langle S_x \rangle = z(ab^* + a^*b).
\end{equation}
Now, due to the angular uncertainty in measurement, and the error in
probabilities due to the preparation procedure, equations
(\ref{valoresp}) and (\ref{probs}), we have:
\begin{eqnarray}
\langle S_x^{\Delta \theta} \rangle &\approx& \langle S_x \rangle +
\Delta \theta \langle S_z \rangle \\
&\approx& z\bigg( \Big(a - \frac{\Delta \theta}{2}b\Big)
\Big(b + \frac{\Delta \theta}{2}a\Big)^* + CC \bigg)
+ \Delta \theta \bigg( \vert a \vert^2
- \vert b \vert^2 + \Delta \theta(ab^* + a^*b) \bigg).\nonumber
\end{eqnarray}
As can be seen, there is a factor proportional to $z$, which decreases
with the size of the environment, and then an error of order $(\Delta
\theta)^2$. Given the exponential dependence on $N$ of the coherences,
they will clearly be smaller than the error for environments large
enough. Therefore, we see that collapse cannot be distinguished from
unitary evolution with local observables either.

\section{Conclusions}

We have shown that fundamental quantum noise in the preparation
procedure and in the observables being measured prevents one from
distinguishing between a collapsed state and an evolved state. This is
done by noticing that even if one takes the measuring apparatus to
occupy the whole universe, which would decrease its errors to a
minimum, quantum uncertainties will completely blur the different
outcomes. We show that undecidability ultimately occurs, even though
one cannot define sharply when it occurs.

The proof consists of two parts. First, completing our previous results
to show that global observables cannot be used to check whether
collapse has occurred or not, from a fundamental point of view rather
than with FAPP arguments. Second, we see that similar arguments also
show that local observables cannot be used either, also for
fundamental reasons.

The fact that undecidability can be established in a sharp way and not
``for all practical purposes'' only, allows to construct a realist
interpretation of quantum mechanics based on the definition of event
introduced. This offers further support for the ``Montevideo''
interpretation of quantum mechanics we outlined\cite{montevideo} in a
previous paper.

\section*{Acknowledgments}

We wish to thank \v{C}aslav Brukner and Johannes Kofler for sharing
their results prior to publication.  This work was supported in part
by grant NSF-PHY-0650715, funds of the Hearne Institute for
Theoretical Physics, FQXi, CCT-LSU, Pedeciba and ANII PDT63/076.


\begin{thebibliography}{0}
\bibitem{montevideo}
  R.~Gambini, L.~P.~Garcia-Pintos and J.~Pullin,
  ``Complete quantum mechanics: an axiomatic formulation of the Montevideo
  interpretation,''
  arXiv:1002.4209 [quant-ph];  R.~Gambini and J.~Pullin,
  %``The Montevideo interpretation of quantum mechanics: frequently asked
  %questions,''
  J.\ Phys.\ Conf.\ Ser.\  {\bf 174}, 012003 (2009)
  [arXiv:0905.4402 [quant-ph]];
  R.~Gambini and J.~Pullin,
  ``Free will, undecidability, and the problem of time in quantum gravity,''
  arXiv:0903.1859 [quant-ph].


\bibitem{undecidability}
  R.~Gambini, L.~P.~G.~Pintos and J.~Pullin,
  %``Undecidability and the problem of outcomes in quantum measurements,''
  Found.\ Phys.\  {\bf 40}, 93 (2010)
  [arXiv:0905.4222 [quant-ph]].


\bibitem{zurek} W. Zurek, Phys. Rev. {\bf D26}, 1862 (1982).

\bibitem{despagnat}
B. d'Espagnat ``Veiled reality'', Addison Wesley,
  New York (1995).

\bibitem{obregon}
  R.~Gambini, R.~Porto and J.~Pullin,
  %``Fundamental decoherence from quantum gravity:
  %A pedagogical review,''
  Gen.\ Rel.\ Grav.\  {\bf 39}, 1143 (2007).

\bibitem{ng}
F. K\'arolhy\'azy, A. Frenkel, B. Luk\'acs in ``Quantum concepts in
space and time'' R. Penrose and C. Isham, editors, Oxford University
Press, Oxford (1986); Y.~J.~Ng and H.~van Dam, Annals N.\ Y.\ Acad.\
Sci.\ {\bf 755}, 579 (1995) [arXiv:hep-th/9406110]; Mod.\ Phys.\
Lett.\ A {\bf 9}, 335 (1994); G.~Amelino-Camelia, %``Limits On The
Measurability Of Space-Time Distances In The Semiclassical
%Approximation Of Quantum Gravity,'' Mod.\ Phys.\ Lett.\ A {\bf 9},
3415 (1994) [arXiv:gr-qc/9603014]; S. Lloyd, J. Ng, Scientific
American, November (2004).


\bibitem{koflerbrukner} C. Brukner, J. Kofler, ``Are there fundamental
  limits for observing quantum phenomena from within quantum
  theory?'', arXiv:1009.2654 [quan-ph]
\end{thebibliography}
\end{document}